\documentclass[%
 reprint,
 amsmath,amssymb,
 aps,
]{revtex4-2}

\usepackage{graphicx}
\usepackage{dcolumn}
\usepackage{bm}

\begin{document}

\preprint{APS/123-QED}

\title{Quantum Gaussian process regression}
\author{Meng-han Chen$^{1}$}
\author{Song Lin$^{2}$}
\altaffiliation{Corresponding author. Email address: lins95@gmail.com}
\author{Gong-De Guo$^{1}$}
\author{Jing Li$^{1}$}
\affiliation{
	$^{1}$College of Mathematics and Informatics, Fujian Normal University, Fuzhou 350117, China\\
	$^{2}$Digital Fujian Internet-of-Things Laboratory of Environmental Monitoring, Fujian Normal University, Fuzhou 350117, China}
\date{\today}

\begin{abstract}
In this paper, a quantum algorithm based on gaussian process regression model is proposed. The proposed quantum algorithm consists of three sub-algorithms. One is the first quantum sub-algorithm to efficiently generate mean predictor. The improved HHL algorithm is proposed to obtain the sign of outcomes. Therefore, the terrible situation that results is ambiguous in terms of original HHL algorithm is avoided, which makes whole algorithm more clear and exact. The other is to product covariance predictor with same method. Thirdly, the squared exponential covariance matrices are prepared that annihilation operator and generation operator are simulated by the unitary linear decomposition Hamiltonian simulation and kernel function vectors is generated with blocking coding techniques on covariance matrices. In addition, it is shown that the proposed quantum gaussian process regression algorithm can achieve quadratic faster over the classical counterpart.
\end{abstract}
 
\pacs{03.67.Dd, 03.65.Ta, 03.67.Hk}
\keywords{Quantum gaussian process regression, non-sparse hamiltonian simulation, harmonic oscillator, quadratic acceleration} 

\maketitle


\section{Introduction}

As a subfield at the intersection of computer science and statistics, building a system known as machine learning that is adaptive and able to learn from experience, which has attracted researchers from many fields. However, with the rapid development of information technology, the total amount of global data grows exponentially every year, which makes classical machine learning algorithm will face great challenges in computing performance when processing big data in the future. Quantum computing uses the basic characteristics of quantum mechanics (such as quantum superposition and quantum entanglement) to achieve computing tasks and has a significant speed advantage over classical computing in solving some specific problems \cite{BWP17,AT17}. For example, Shor's quantum factoring algorithm has an exponential acceleration over the classical algorithm  \cite{SS97}, posing a serious threat to the security of RSA cryptographic system which is widely used. In recent years, quantum computing has been applied to the field of machine learning. A variety of efficient quantum machine learning algorithms have been proposed, such as quantum clustering analysis \cite{C15,YGC16}, quantum neural network \cite{CLM19,FN18}, quantum classification \cite{ZAEFAEA19,S20}, quantum decision tree \cite{S02}, quantum association rules \cite{YGW16} and so on. The researches of quantum algorithm and the exploration of the realization method of quantum mechanical properties in the algorithm field can make the further development of artificial intelligence algorithms \cite{SF18}. Therefore, it is significantly important to combine quantum information with machine learning to come up with more efficient quantum algorithms \cite{LD98}.

One of most positive directions in machine learning is the development of practical Bayesian approaches to solve real challenging problems \cite{H20}. Among them, gaussian process is one of most important Bayesian machine learning methods, which is based on a particularly efficient method to place a priori function on the function space. Gaussian process regression was proposed by Carl E. Rasmussen and scholar Christopher K. I. Williams in 1995 \cite{WR95}, aiming at using gaussian process priori to the process regression analysis of data. The model hypothesis includes noise (regression residual) and gaussian process priori. Its solution is carried out according to bayesian inference without limiting the form of its kernel function. It can be used in the fields of time series analysis, image processing and automatic control \cite{DPSC15,LPL13} in terms of convenience. Therefore, this paper attempts to combine gaussian process regression with quantum information and proposes the quantum gaussian process regression.

The rest of this paper is organized as follows. The second section is a review of classical gaussian process regression. The third section introduces quantum gaussian process. The time complexity analysis of quantum algorithm is given in the fourth chapter, The article is ended in the fifth  section with a concluding remark.

\section{\label{sec:level2}A review of classical gaussian process regression}

In the simple regression model, a training dataset $D$ with $\mathit{M}$ data points $\left( {\vec x_i ,y_i } \right)_{i = 1}^M$ is given, where $\vec x_i  = \left( {x_{i1} ,x_{i2} ,...,x_{iN} } \right)^T  \in R^N$ is a column vector of independent input variables and $y_i$ is the corresponding scalar of single output variable. The goal of gaussian process regression is to train a linear function $y = f\left( {\vec x} \right) + \varepsilon$ in a limited set of training dataset $D$ that can fit the relationship between $\vec x_i $ and $y_i$ as well as possible, where $f\left( {\vec x} \right)$ is real value, $y$ is desired value and $\varepsilon  \sim {\rm N}\left( {0,\sigma _M^2 } \right)$ is identically distributed gaussian noise which is independent. Once obtained, $f$ can be used to predict the output $f\left( {\vec x_ *  } \right)$ of a new input data ${\vec x_ *  }$.

A gaussian process regression model is fully specified by a mean function $m\left( {\vec x} \right)$ and a covariance function (also known as a kernel) $k\left( {\vec x,\vec x'} \right)$. In 2005, Rasmussen and Williams \cite{WR05} pointed that these two functions can be obtained in terms of weight-space view and function-space view, expressions are as follows,
\begin{equation}
m\left( {\vec x} \right) = E\left[ {f\left( {\vec x} \right)} \right],
\label{eq:1}
\end{equation}
\begin{equation}
k\left( {\vec x,\vec x'} \right) = E\left[ {\left( {f\left( {\vec x} \right) - m\left( {\vec x} \right)} \right)\left( {f\left( {\vec x'} \right) - m\left( {\vec x'} \right)} \right)} \right],
\label{eq:2}
\end{equation}
where $E$ denotes expectation value. Thus, gaussian process regression can be written as $f\left( {\vec x} \right) \sim gp\left( {m\left( {\vec x} \right),k\left( {\vec x,\vec x'} \right)} \right)$. Here, $gp$ means gaussian process. Generally, classical gaussian process regression specified covariance function as squared exponential covariance function \cite{WR05}, which is denoted as 
\begin{equation}
{\mathop{\rm cov}} \left( {f\left( {\vec x_p ,\vec x_q } \right)} \right) = k\left( {\vec x_p ,\vec x_q } \right) = \exp \left( { - {\textstyle{1 \over 2}}\left| {\vec x_p  - \vec x_q } \right|^2 } \right).
\label{eq:3}
\end{equation}
Therefore, the central goal in gaussian process regression models is to predict the mean value and the covariance value of this distribution, also known as mean predictor ${{\bar f}}_{\rm{*}}$ and variance predictor $V\left[ {f_* } \right]$. The deduction is presented in Ref.\cite{WR05}, through deduction, mean predictor and variance predictor can be re-expressed as  
\begin{equation}
{{\bar f}}_{{*}}=  {{ \vec k}}_*^T \left( {K + \sigma _M^2 I} \right)^{ - 1} \vec y,
\label{eq:4}
\end{equation}
\begin{equation}
V\left[ {f_* } \right] = k\left( {x_* ,x_* } \right) - {{\vec k}}_*^T \left( {K + \sigma _M^2 I} \right)^{ - 1} {{\vec k}}_*,
\label{eq:5}
\end{equation}
where ${{\vec k}}_*$ is a vector that represents the kernel function of a test point $\vec x_* $ and all training points ${\vec x}$'s; $K$ denotes a covariance matrix of $M \times M$ dimension, that is calculation between $M$ training datas; $k\left( {x_* ,x_* } \right)$ is covariance of test point $\vec x_* $ and itself, which is a constant. Taking Eq. (4) as a linear combination of $M$ kernel functions, each of them is centered on training point, thus, Eq. (4) can be written as
\begin{equation}
\bar f\left( {\vec x_* } \right) = \sum\limits_{i = 1}^M {\alpha _i k\left( {\vec x_i ,\vec x_* } \right)},
\label{eq:6}
\end{equation}
here $\vec \alpha  = \left( {K + \sigma _M^2 I} \right)^{ - 1} {{\vec y}}$,

Then, we introduce the realization process of gaussian process distribution in classical calculation. Firstly, calculating the Cholesky decomposition of $L: = cholesky\left( {K + \sigma _M^2 I} \right)$, thus $\alpha  = L^T {\rm{\backslash }}\left( {{\rm{L\backslash y}}} \right)$. Finally,  mean predictor can be computed by $\bar f_*  = {{\vec k}}_{{*}}^T \vec \alpha$. Since the calculation of  Cholesky factor is numerically stable, runtime is proportional to $O\left( {M^3 } \right)$. For $V\left[ {f_* } \right]$, let $\vec v: = L{{\backslash \vec k}}_{\rm{*}}$, thus $V\left[ {f_* } \right]: = k\left( {x_* ,x_* } \right) - {\vec v}^T \vec v$ can be computed with a number of basic arithmetrics. The computation of $k\left( {x_* ,x_* } \right)$ requires only the time of constant term. Therefore, total runtime is $O\left( {M^3 } \right)$. For the current era of big data, the number of input points is huge, consequently, time complexity is quite large. That's why we want to use quantum information.

\section{Quantum Gaussian process regression}
Observing mean predictor and variance predictor, there is an inverse process. Thus, it is easy to think of using the HHL algorithm \cite{HHL09} as a subroutine to get desired result. Firstly, covariance matrix $K$ is a real covariance matrix, hence, $K$ is a real symmetric matrix, namely $K$ is Hermitian matrix. Writing $K$ in the spectral decomposition form \cite{OSA99}
\begin{equation}
K = \sum\limits_{j = 0}^{M - 1} {\lambda _j \left| {u_j } \right\rangle \left\langle {u_j } \right|},
\label{eq:7}
\end{equation} 
where $\left\{ {\lambda _j } \right\}_{j = 0}^{M - 1}$ are the eigenvalues of $K$ and $\left\{ {\left| {u_j } \right\rangle } \right\}_{j = 0}^{M{\rm{ - }}1}$ are the corresponding eigenvector of $\left\{ {\lambda _j } \right\}_{j = 0}^{M - 1}$. ${{{{\vec k}}_* } \mathord{\left/{\vphantom {{{{\vec k}}_* } {\left\| {{{\vec k}}_* } \right\|}}} \right.\kern-\nulldelimiterspace} {\left\| {{{\vec k}}_* } \right\|}}$ can be represented the linear combination of $\left\{ {\left| {u_j } \right\rangle } \right\}_{j = 0}^{M{\rm{ - }}1}$, that is ${{{{\vec k}}_* } \mathord{\left/{\vphantom {{{{\vec k}}_* } {\left\| {{{\vec k}}_* } \right\|}}} \right.\kern-\nulldelimiterspace} {\left\| {{{\vec k}}_* } \right\|}} = \sum\limits_{j = 0}^{M - 1} {\alpha _j \left| {u_j } \right\rangle _1 }$. Similarly, ${{\vec y} \mathord{\left/{\vphantom {{\vec y} {\left\| {\vec y} \right\|}}} \right.
\kern-\nulldelimiterspace} {\left\| {\vec y} \right\|}}$ can also be represented by $\left\{ {\left| {u_j } \right\rangle } \right\}_{j = 0}^{M{\rm{ - }}1}$, namely ${{\vec y} \mathord{\left/
{\vphantom {{\vec y} {\left\| {\vec y} \right\|}}} \right.\kern-\nulldelimiterspace} {\left\| {\vec y} \right\|}} = \sum\limits_{j = 0}^{M - 1} {\beta _j \left| {u_j } \right\rangle _1 }$. Therefore, the expressions of Eq. (4) and Eq. (5) in the mathematical context are  
\begin{equation}
\bar f_{\rm{*}} {\rm{ = }}\frac{{\alpha _{\rm{j}} \beta _j }}{{\lambda _{\rm{j}}  + \sigma _M^2 }}\left\| {\vec k_ *  } \right\|\left\| {\vec y} \right\|,
\label{eq:8}
\end{equation}
\begin{equation}
V\left[ {f_* } \right] = k\left( {x_* ,x_* } \right) - \frac{{\alpha _j^2 }}{{\lambda _j {\rm{ + }}\sigma _M^2 }}\left\| {\vec k_ *  } \right\|^2.
\label{eq:9}
\end{equation}
Next, how to get the values of Eq. (8) and Eq. (9) is described by the following quantum steps.

\subsection{Preparing quantum state of ${{\vec y}}$ and ${{\vec k}}_* $}
The specific expression is $\left| {{{\vec y}}} \right\rangle _{\rm{1}} {\rm{ = }}\sum\limits_{j = 0}^{M{\rm{ - }}1} {\beta _j \left| {\vec u_j } \right\rangle _1 }$ and $\left| {{{\vec k}}_* } \right\rangle _{\rm{1}} {\rm{ = }}\sum\limits_{j = 0}^{M{\rm{ - }}1} {\alpha _j \left| {\vec u_j } \right\rangle _1 }$. Ref. \cite{YGW21} pointed that the procedure of preparing quantum state $\left| {{{\vec y}}} \right\rangle _{\rm{1}}$ and $\left| {{{\vec k}}_* } \right\rangle _{\rm{1}}$
is as follows. Firstly, supposing that the quantum oracles $O_{\vec y}$ and $O_{\vec k_* }$ are provided, which can efficiently access the entries of $\vec y$ and $\vec k_*$ in time $O\left( {{\text{poly}}\log M} \right)$ from quantum random access memory (QRAM) and act as $O_{\vec y} :\left| j \right\rangle _1 \left| 0 \right\rangle _{a1}  \mapsto \left| j \right\rangle _1 \left| {y_j } \right\rangle _{a1}$ and $O_{\vec k_* } :\left| j \right\rangle _1 \left| 0 \right\rangle _{b1}  \mapsto \left| j \right\rangle _1 \left| {k_{*j} } \right\rangle _{b1}$. Secondly, giving initial state $\left| 0 \right\rangle _1 \left| 0 \right\rangle _{a1}$ and performing the quantum Fourier transform on the first register, which gets $\sum\limits_{j = 0}^{M - 1} {\frac{1}{{\sqrt M }}\left| j \right\rangle _1 } \left| 0 \right\rangle _{a1}$. The quantum Fourier transform on an orthonormal basis $\left| 0 \right\rangle ,...,\left| {N - 1} \right\rangle $ is defined to be a linear operator with following action on the basis states, 
\begin{equation}
\left| j \right\rangle  \to \frac{1}{{\sqrt N }}\sum\limits_{k = 0}^{N - 1} {{e^{{{2\pi ijk} \mathord{\left/
					{\vphantom {{2\pi ijk} N}} \right.
					\kern-\nulldelimiterspace} N}}}\left| k \right\rangle } .
\end{equation}
Thirdly, applying the oracle $O_{\vec y}$ on $\sum\limits_{j = 0}^{M - 1} {\frac{1}{{\sqrt M }}\left| j \right\rangle _1 } \left| 0 \right\rangle _{a1}$, which obtains $
\sum\limits_{j = 0}^{M - 1} {\frac{1}{{\sqrt M }}\left| j \right\rangle _1 } \left| {y_j } \right\rangle _{a1}\textbf{}$. Then, appending one qubit and performing conditioned rotation on the register a1, which gains state 
\begin{equation}
\sum\limits_{j = 0}^{M - 1} {\frac{1}{{\sqrt M }}\left| j \right\rangle _1 } \left| {y_j } \right\rangle _{a1} \left( {\sqrt {1{\rm{ - }}\left( {\frac{{y_j }}{{\left\| {{{\vec y}}} \right\|_{\max } }}} \right)^2 } \left| 0 \right\rangle _{{\rm{a2}}} {\rm{ + }}\frac{{y_j }}{{\left\| {{{\vec y}}} \right\|_{\max } }}\left| 1 \right\rangle _{a2} } \right).
\end{equation} 
Finally, performing inverse oracle operation and measuring the last register in the basis $
\left\{ {\left| 0 \right\rangle ,\left| 1 \right\rangle } \right\}$, the remainder particles are in the state $\frac{{\sum\limits_{j = 0}^{M - 1} {y_j \left| j \right\rangle _1 } }}{{\left\| {\vec y} \right\|_{\max } }}$, which has probability  $p_{{y}}  = \frac{{\sum\limits_{j = 0}^{M - 1} {y_j^2 } }}{{M\left\| {{{\vec y}}} \right\|_{\max }^2 }}{\rm{ = }}\Omega \left( 1 \right)$ ($\vec y$ and ${{\vec k}}_* $ are balanced). Thus, we need to perform $O\left( 1 \right)$ to get $\left| {\vec y} \right\rangle$. Moreover, $\left\| {\vec y} \right\|^2  = \sum\limits_{j = 0}^{M - 1} {y_j^2 }  = MP_y \left\| {{{\vec y}}} \right\|_{\max }^2$  can also be estimated. The total runtime of above operation is $O\left( {{\text{poly}}\log M} \right)$. In a similar way,  $\left| {{{\vec k}}_* } \right\rangle$ and $\left\| {{{\vec k}}_* } \right\|^2$ can be obtained in time $O\left( {{\text{poly}}\log M} \right)$.

\subsection{\label{sec:citeref}Preparing $U$ operator that makes $U\left| {{{\vec k}}_* } \right\rangle \left| 0 \right\rangle  = \left| {\varphi _3 }\right\rangle$}
Step B1 Assuming that covariance matrix $K$ is prepared (The detailed preparation method will be described in section E). The matrix $K$ is Hermitian but not necessarily sparse. Thus, the unitary linear decomposition (LCU) approach \cite{BCCKD15} is adopted to take the exponentiation of matrix. LCU means under the assumptions that the quantum oracle $O_{\vec k} \left| 0 \right\rangle ^{ \otimes \left\lceil {\log M} \right\rceil }  = \sum\nolimits_{i = 0}^{M - 1} {\sqrt {k_i } } \left| i \right\rangle$, which can be efficiently implemented in time $O\left( {{\rm{poly}}\log M} \right)$ and $\sum\nolimits_{i = 0}^{M - 1} {k_i  = 1}\textbf{}$ is provided. Zhou and Wang \cite{ZW17} proposed that $e^{ - iKt}$ can be simulated within spectral-norm error $\varepsilon$ in time $O\left( {\frac{{t{\text{poly}}\log M\log \left( {\frac{t}
				{\varepsilon }} \right)}}
	{{\log \log \left( {\frac{t}
				{\varepsilon }} \right)}}} \right)$. Noted that the algorithm decomposes $K$ into $M$ linear combinations of unitary operators that can be efficiently implemented, that is $K = \sum\nolimits_{j = 0}^{M - 1} {k_j V_j }$, where $V_j  = \sum\nolimits_{l = 0}^{M - 1} {\left| {\left( {l - j + 1} \right)\bmod \left| M \right\rangle \left\langle l \right|} \right|} ,j = 1,2,...,M$ can be implemented using $O\left( {\log M} \right)$ one- or two-qubit gates. 

Step B2 Performing the phase estimation \cite{DGH18} on $\left| {{{\vec k}}_* } \right\rangle _1  = \sum\limits_{j = 0}^{M{\rm{ - }}1} {\alpha _j \left| {u_j } \right\rangle _1 }$, we can obtain 
\begin{equation}
\left| {\varphi _1 } \right\rangle _{{\rm{12}}} {\rm{ = }}\sum\limits_{j = 0}^{M{\rm{ - }}1} {\alpha _j \left| {u_j } \right\rangle _1 } \left| {\lambda _j } \right\rangle _2.
\label{eq:10}
\end{equation}

Step B3 Adding an ancilla qubit $\left| 0 \right\rangle _3$ and performing controlled rotation on it controlled on the second register, the state becomes
\begin{equation}
\begin{array}{l}
\left| {\varphi _2 } \right\rangle _{{\rm{123}}} {\rm{ = }}\sum\limits_{j = 0}^{M{\rm{ - }}1} {\alpha _j \left| {u_j } \right\rangle _1 } \left| {\lambda _j } \right\rangle _2  \\ 
\left( {\sqrt {1{\rm{ - }}\left( {\frac{c}{{\lambda _j  + \sigma _M^2 }}} \right)^2 } \left| 0 \right\rangle _3  + \frac{c}{{\lambda _j  + \sigma _M^2 }}\left| 1 \right\rangle _3 } \right) \\ 
\end{array}
\label{eq:11}
\end{equation}

Step B4 Performing inverse phase estimation and discarding the second register, we can get 
\begin{equation}
\begin{array}{l}
\left| {\varphi _3 } \right\rangle _{{\rm{13}}} {\rm{ = }}\sum\limits_{j = 0}^{M{\rm{ - }}1} {\alpha _j \left| {u_j } \right\rangle _1 }  \\ 
\left( {\sqrt {1{\rm{ - }}\left( {\frac{c}{{\lambda _j  + \sigma _M^2 }}} \right)^2 } \left| 0 \right\rangle _3  + \frac{c}{{\lambda _j  + \sigma _M^2 }}\left| 1 \right\rangle _3 } \right) \\ 
\end{array}.
\label{eq:12}
\end{equation}
The above operations are all unitary operations, thus, above processes can be denoted with $U$.

\subsection{Computing predictive mean value in the quantum context}
Step C1 Preparing quantum initial state $\left| 0 \right\rangle _1 \left| 0 \right\rangle _2 \left( {\left| 0 \right\rangle _3 \left| {{{\vec k}}_* } \right\rangle _4  + \left| 1 \right\rangle _3 \left| {{{\vec y}}} \right\rangle _4 } \right)\left| 0 \right\rangle _5$. When the third register is  $\left| 0 \right\rangle _3$, unitary operation $U$ on the fourth and fifth register is executed. Otherwise,  pauli-X on the fifth register is performed. Thus, the state of system is  
\begin{equation}
\left| 0 \right\rangle _1 \left| 0 \right\rangle _2 \left( {\left| 0 \right\rangle _3 \left| {\varphi _3 } \right\rangle _{45}  + \left| 1 \right\rangle _3 \left| {{{\vec y}}} \right\rangle _4 \left| 1 \right\rangle _5 } \right).
\end{equation}

Step C2 Performing the Hadamard transform on the first register. In the meanwhile, executing pauli-X on the second register. Then, the Hadamard transform on the second register is carried out. In this way, the state becomes
\begin{equation}
\frac{1}{2}\left( {\left| 0 \right\rangle _{\rm{1}} {\rm{ + }}\left| 1 \right\rangle _1 } \right)\left( {\left| 0 \right\rangle _{\rm{2}} {\rm{ - }}\left| 1 \right\rangle _2 } \right)\left( {\left| 0 \right\rangle _3 \left| {\varphi _3 } \right\rangle _{45}  + \left| 1 \right\rangle _3 \left| {{{\vec y}}} \right\rangle _4 \left| 1 \right\rangle _5 } \right).
\end{equation}

Step C3 Doing nothing when the first register is $\left| 0 \right\rangle _1$. Otherwise, performing swap operation \cite{BCW01} on the second and the third register, which obtains  
\begin{equation}
\begin{gathered}
\frac{1}
{2}\left[ {\left| 0 \right\rangle _1 \left( {\left| 0 \right\rangle _2 \left| 0 \right\rangle _3 \left| {\varphi _3 } \right\rangle _{45} {\text{ + }}\left| 0 \right\rangle _2 \left| 1 \right\rangle _3 \left| {\vec y} \right\rangle _4 \left| 1 \right\rangle _5 } \right.} \right. \hfill \\
{\text{            }}\left. { - \left| 1 \right\rangle _2 \left| 0 \right\rangle _3 \left| {\varphi _3 } \right\rangle _{45}  - \left| 1 \right\rangle _2 \left| 1 \right\rangle _3 \left| {\vec y} \right\rangle _4 \left| 1 \right\rangle _5 } \right) \hfill \\
{\text{     + }}\left| 1 \right\rangle _1 \left( {\left| 0 \right\rangle _2 \left| 0 \right\rangle _3 \left| {\varphi _3 } \right\rangle _{45} {\text{ + }}\left| 1 \right\rangle _2 \left| 0 \right\rangle _3 \left| {\vec y} \right\rangle _4 \left| 1 \right\rangle _5 } \right. \hfill \\
{\text{           }}\left. {\left. { - \left| 0 \right\rangle _2 \left| 1 \right\rangle _3 \left| {\varphi _3 } \right\rangle _{45}  - \left| 1 \right\rangle _2 \left| 1 \right\rangle _3 \left| {\vec y} \right\rangle _4 \left| 1 \right\rangle _5 } \right)} \right]. \hfill \\ 
\end{gathered} 
\end{equation}

Step C4 Measuring the first register of Eq. (17) with operator $M = \frac{1}{2}\left( {\left| 0 \right\rangle  - \left| 1 \right\rangle } \right)\left( {\left\langle 0 \right| - \left\langle 1 \right|} \right)$, where the inner product of Eq. (14) and $\left| {\vec y} \right\rangle _4 \left| 1 \right\rangle _5 $ is $\sum\limits_{j = 0}^{M - 1} {\frac{{c\alpha _j \beta _j }}{{\lambda _j  + \sigma _M^2 }}}$. Thus, the probability of measurement is $p = \frac{1}{2}\left( {1 + \sum\limits_{j = 0}^{M - 1} {\frac{{c\alpha _j \beta _j }}{{\lambda _j  + \sigma _M^2 }}} } \right)$. Here, $c$ is a constant. Through measurement, the specific value of $\sum\limits_{j = 0}^{M - 1} {\frac{{c\alpha _j \beta _j }}{{\lambda _j  + \sigma _M^2 }}}$ is attained. Therefore, predicting mean value in quantum context is achieved. 

\subsection{Computing predictive covariance value in the quantum context}
Step D1 Preparing quantum initial state $\left| 0 \right\rangle _1 \left| 0 \right\rangle _2 \left( {\left| 0 \right\rangle _3 \left| {{{\vec k}}_* } \right\rangle _4  + \left| 1 \right\rangle _3 \left| {{{\vec k}}_* } \right\rangle _4 } \right)\left| 0 \right\rangle _5$. When the third register is in the state $\left| 0 \right\rangle _3$, unitary operation $U$ on the fourth register is performed. Otherwise, pauli-X operation is performed on the fifth register. Thus, we can get $\left| 0 \right\rangle _1 \left| 0 \right\rangle _2 \left( {\left| 0 \right\rangle _3 \left| {\varphi _3 } \right\rangle _{45}  + \left| 1 \right\rangle _3 \left| {{{\vec k}}_* } \right\rangle _4 \left| 1 \right\rangle _5 } \right)$.

Step D2 Performing the Hadamard transform on the first register. At the same time, executing pauli-X on the second register. Then the Hadamard transform on the second register is carried out, which results in $\frac{1}{2}\left( {\left| 0 \right\rangle _{\rm{1}} {\rm{ + }}\left| 1 \right\rangle _1 } \right)\left( {\left| 0 \right\rangle _{\rm{2}} {\rm{ - }}\left| 1 \right\rangle _2 } \right)\left( {\left| 0 \right\rangle _3 \left| {\varphi _3 } \right\rangle _{45}  + \left| 1 \right\rangle _3 \left| {{{\vec k}}_* } \right\rangle _4 \left| 1 \right\rangle _5 } \right)$.

Step D3 Doing nothing when the first register is $\left| 0 \right\rangle _1$. Otherwise, performing swap operation for the second and the third register. Thus, the state can be obtained 
\begin{equation}
\begin{gathered}
\frac{1}
{2}\left[ {\left| 0 \right\rangle _1 \left( {\left| 0 \right\rangle _2 \left| 0 \right\rangle _3 \left| {\varphi _3 } \right\rangle _{45} {\text{ + }}\left| 0 \right\rangle _2 \left| 1 \right\rangle _3 \left| {\vec k_* } \right\rangle _4 \left| 1 \right\rangle _5 } \right.} \right. \hfill \\
{\text{            }}\left. { - \left| 1 \right\rangle _2 \left| 0 \right\rangle _3 \left| {\varphi _3 } \right\rangle _{45}  - \left| 1 \right\rangle _2 \left| 1 \right\rangle _3 \left| {\vec k_* } \right\rangle _4 \left| 1 \right\rangle _5 } \right) \hfill \\
{\text{     + }}\left| 1 \right\rangle _1 \left( {\left| 0 \right\rangle _2 \left| 0 \right\rangle _3 \left| {\varphi _3 } \right\rangle _{45} {\text{ + }}\left| 1 \right\rangle _2 \left| 0 \right\rangle _3 \left| {\vec k_* } \right\rangle _4 \left| 1 \right\rangle _5 } \right. \hfill \\
{\text{           }}\left. {\left. { - \left| 0 \right\rangle _2 \left| 1 \right\rangle _3 \left| {\varphi _3 } \right\rangle _{45}  - \left| 1 \right\rangle _2 \left| 1 \right\rangle _3 \left| {\vec k_* } \right\rangle _4 \left| 1 \right\rangle _5 } \right)} \right]. \hfill \\ 
\end{gathered} 
\end{equation}

Step D4 Measuring the first register with operator $M = \frac{1}{2}\left( {\left| 0 \right\rangle  - \left| 1 \right\rangle } \right)\left( {\left\langle 0 \right| - \left\langle 1 \right|} \right)$, where the inner product of Eq. (14) and $\left| {{{\vec k}}_* } \right\rangle _4 \left| 1 \right\rangle _5$ is $\sum\limits_{j = 0}^{M - 1} {\frac{{c'\alpha _j^2 }}{{\lambda _j  + \sigma _M^2 }}}$. Thus, the probability of measurement is $p = \frac{1}{2}\left( {1 + \sum\limits_{j = 0}^{M - 1} {\frac{{c'\alpha _j^2 }}{{\lambda _j  + \sigma _M^2 }}} } \right)$. Here, $c’$ is a constant. We can get specific value of $\sum\limits_{j = 0}^{M - 1} {\frac{{c'\alpha _j^2 }}{{\lambda _j  + \sigma _M^2 }}}$ from probability. Therefore, predicting covariance value in quantum context is achieved. 

\subsection{Preparing covariance matrix $K$ and kernel function vector $\vec k_*$}
In classical gaussian process regression, covariance function is generally ${\mathop{\rm cov}} \left( {f\left( {\vec x_p ,\vec x_q } \right)} \right) = k\left( {\vec x_p ,\vec x_q } \right) = \exp \left( { - {\textstyle{1 \over 2}}\left| {\vec x_p  - \vec x_q } \right|^2 } \right)$. Thus, the representation of covariance matrix and kernel function vector are respectively 
\begin{equation}
K = \sum\limits_{p = 0}^{M - 1} {\sum\limits_{q = 0}^{M - 1} {\exp \left( { - \frac{1}
			{2}\left| {\vec x_p  - \vec x_q } \right|^2 } \right)} } \left| p \right\rangle \left\langle q \right|
\end{equation}
\begin{equation}
\vec k_*  = \sum\limits_{p = 0}^{M - 1} {\exp \left( { - \frac{1}
		{2}\left| {\vec x_*  - \vec x_p } \right|^2 } \right)} \left| p \right\rangle.
\end{equation}

The preparation of quantum initial states has always been difficult in quantum machine learning. And in this paper, it is also a challenge. Inspired by quantum radial basis network \cite{S20}, coherent state \cite{S12,F01} and block coding techniques \cite{S19,CGJ19} are utilized to achieve the preparation of initial states. Coherent states play a crucial role in quantum optics and mathematical physics. They are defined in the Fock states $\left\{ {\left| 0 \right\rangle ,\left| 1 \right\rangle ,...} \right\}$, which is a basis of the infinitely dimensional Hilbert space ${\rm H}$. Let’s $a,a^\dag  $  respectively, be the annihilation and creation operators of the harmonic oscillator. Then 
\begin{equation}
a\left| n \right\rangle  = \sqrt n \left| {n - 1} \right\rangle, 
a^\dag  \left| n \right\rangle  = \sqrt {n + 1} \left| {n + 1} \right\rangle.
\label{eq:13}
\end{equation}

For any $n \ge 1$, it is easy to see that
\begin{equation}
\left| n \right\rangle  = \frac{{\left( {a^\dag  } \right)^n }}{{\sqrt {n!} }}\left| 0 \right\rangle. 
\label{eq:14}
\end{equation}

Let $r \in R$ be a real number, its coherent state is defined by
\begin{equation}
\left| {\varphi _r } \right\rangle  = e^{ - {{r^2 } \mathord{\left/
			{\vphantom {{r^2 } 2}} \right.
			\kern-\nulldelimiterspace} 2}} \sum\limits_{k = 0}^\infty  {\frac{{r^k }}{{\sqrt {k!} }}} \left| k \right\rangle. 
\label{eq:15}
\end{equation}
It is a unit eigenvector of a corresponding to the eigenvalue $r$, that is $a\left| {\varphi _r } \right\rangle  = r\left| {\varphi _r } \right\rangle$. From Eq. (19) and Eq. (20), we also have $\left| {\varphi _r } \right\rangle  = e^{ - {{r^2 } \mathord{\left/
			{\vphantom {{r^2 } 2}} \right.
			\kern-\nulldelimiterspace} 2}} e^{ra^\dag  } \left| 0 \right\rangle  = e^{r\left( {a^\dag   - \frac{a}{2}} \right)} \left| 0 \right\rangle$, thus $\left| {\varphi _r } \right\rangle $ is obtained by a unitary operator of dimension infinity.
		
As for the preparation of $\left| {\varphi _r } \right\rangle $  in a finite quantum circuit, we can consider its Taylor approximation:
\begin{equation}
\left| {\tilde \varphi _r } \right\rangle  \propto e^{ - {{r^2 } \mathord{\left/
			{\vphantom {{r^2 } 2}} \right.
			\kern-\nulldelimiterspace} 2}} \sum\nolimits_{k = 0}^{T - 1} {\frac{{r^k }}{{\sqrt {k!} }}\left| k \right\rangle }.
\label{eq:16}
\end{equation}
Thus, the upper bound on error square is 
\begin{equation}
\left| {\left| {\varphi _r } \right\rangle  - \left| {\tilde \varphi _r } \right\rangle } \right|^2  \le \frac{{r^{2T} }}{{T!}}.
\label{eq:17}
\end{equation}

By stirling formula, we can get $T! \approx \sqrt {2\pi T} \left( {\frac{T}{e}} \right)^T$. Now, keeping error within $\delta$. Let's take Eq. (17) $\leqslant \delta ^2 $, that is $T$ only need to satisfy  $2\log \frac{1}
{\delta } - \frac{1}
{2}\log 2\pi  \leqslant \left( {T + \frac{1}
	{2}} \right)\log T - 2T\log r - T$. Thus, it is easy to get the Taylor approximation of $\left| {\varphi _r } \right\rangle$. 

From the previous analysis, $\left| {\varphi _r } \right\rangle $ can be obtained by an unitary operator applying to $\left| 0 \right\rangle$. Because annihilation operators and creation operators are Hermitian, operator $e^{r(a^\dag   - \frac{1}{2}a)}$ can be simulated by LCU. It is also shown that $e^{r(a^\dag   - \frac{1}{2}a)}$ can simulate it with finite dimensions, which get a similar result.
Thus for any vector ${{\vec x = }}\left( {x_1 ,x_2 ,...,x_N } \right)^T$, the definition of coherent state is $\left| {\varphi _x } \right\rangle  = \left| {\varphi _{x^{(1)} } } \right\rangle  \otimes \left| {\varphi _{x^{(2)} } } \right\rangle  \otimes  \cdots  \otimes \left| {\varphi _{x^{(N)} } } \right\rangle$, which costs $O\left( {\frac{{Nt{\text{poly}}\log N\log \left( {\frac{t}
				{\varepsilon }} \right)}}
	{{\log \log \left( {\frac{t}
				{\varepsilon }} \right)\delta }}} \right)$. Now, considering the superposition states of training samples' coherent states:
\begin{equation}
\left| \Psi  \right\rangle {\rm{ = }}\frac{1}{{\sqrt M }}\sum\limits_{p = 0}^{M - 1} {\left| p \right\rangle _{e1}} \left| {\varphi _{x_p } } \right\rangle _{e2}.
\label{eq:18}
\end{equation}
Firstly, applying the Fourier transform on $\left| 0 \right\rangle _{e1}$, which results in $\frac{1}
{{\sqrt M }}\sum\limits_{p = 0}^{M - 1} {\left| p \right\rangle _{e1} }$. Then appending one qubit, which gets $\frac{1}{{\sqrt M }}\sum\limits_{p = 0}^{M - 1} {\left| p \right\rangle _{e1} \left| 0 \right\rangle _{e2}}$. Finally, applying unitary opeartors to the e2 register, Eq. (26) can be attained in time $O\left( {\frac{{MNt{\text{poly}}\log N\log \left( {\frac{t}
				{\varepsilon }} \right)}}
	{{\log \log \left( {\frac{t}
				{\varepsilon }} \right)\delta }}} \right)$. 
		
Then, taking the partial trace on the e2 register of $\left| \Psi  \right\rangle \left\langle \Psi  \right|$ gives rise to the density operator of covariance matrix
\begin{equation}
\begin{gathered}
\rho  = Tr_2 \left| \Psi  \right\rangle \left\langle \Psi  \right| \hfill \\
{\text{  }} =
\frac{1}
{M}\sum\limits_{p,q = 0}^{M - 1} {\exp \left( { - \frac{1}
		{2}\left| {\vec x_p  - \vec x_q } \right|^2 } \right)\left| p \right\rangle \left\langle q \right|}. \\ 
\end{gathered} 
\label{eq:19}
\end{equation}

For covariance vector $\sum\limits_{p = 0}^{M - 1} {\exp \left( { - \frac{1}
		{2}\left| {\vec x_*  - \vec x_p } \right|^2 } \right)} \left| p \right\rangle$, we put the target vector ${\vec x_* }$ into the training dataset D, which make dataset become $\left( {\vec x_i ,y_i } \right)_{i = 1}^{M + 1} $. Here ${\vec x_{M + 1} }$ is ${\vec x_* }$ and $y_i  = 0$. Then, computing new covariance matrix $K'$ that $K' = \frac{1}
{{M + 1}}\sum\limits_{p,q = 0}^M {\exp \left( { - \frac{1}
{2}\left| {\vec x_p  - \vec x_q } \right|^2 } \right)} \left| p \right\rangle \left\langle q \right|$. 

Ref. \cite{S19} presented a definition that supposed that $K'$   is an $m$-qubit density operator, $\gamma ,\varepsilon ' \in R^\dag$  and $k \in N$. Then there is a $(m+k)$-qubit unitary $U'$ is $\left( {\gamma ,k,\varepsilon '} \right)$ -block-encoding of $A$ so that $\left\| {A  - \gamma \left( {\left\langle 0 \right|^{ \otimes k}  \otimes I_{2m} } \right)U'\left( {\left| 0 \right\rangle ^{ \otimes k}  \otimes I_{2m} } \right)} \right\| \le \varepsilon '$, where $\left\|  \cdot  \right\|$ is the spectral norm, thus, $U '= \left( {\begin{array}{*{20}c}
	{A'} &  \cdot   \\
	\cdot  &  \cdot   \\
	\end{array}} \right)$  satisfied $\left\| {A'  - \gamma A} \right\| \le \varepsilon$. Therefore, $A' \left| q \right\rangle$ can be obtained by considering $U'\left| 0 \right\rangle \left| q \right\rangle$. That is $
U'\left| 0 \right\rangle \left| q \right\rangle  \approx \left( {\begin{array}{*{20}c}
	A'  & .  \\
	. & .  \\
	\end{array}} \right)\left( {\begin{array}{*{20}c}
	{\left| q \right\rangle }  \\
	0  \\
	\end{array}} \right) = \left( {\begin{array}{*{20}c}
	{A' \left| q \right\rangle }  \\
	.  \\
	\end{array}} \right) = A' \left| q \right\rangle \left| 0 \right\rangle  + \left| 0 \right\rangle ^ \bot$. Then, a projection measurement is performed in the basis $\left\{ {\left| 0 \right\rangle ,\left| 1 \right\rangle } \right\}$,  the probability of obtaining measurement outcome $A\left| q \right\rangle $ is $tr\left( {A'^2 } \right)$. Here $A$ is a density operator, thus $tr\left( {A^2 } \right) \leqslant 1$. If $A$ is a pure state, $tr\left( {A^2 } \right) = 1$. Otherwise, $tr\left( {A^2 } \right) < 1$. When the eigenvalues of $A$ are small, we put a coefficient in front of the matrix $A$ in order to obtain $A\left| q \right\rangle $ with a high probability.
Here Gilyén et.al \cite{GSLW18} proposed that the block-encoding of $A$ is $\left( {G^\dag   \otimes I} \right)\left( {I \otimes SWAP} \right)\left( {G \otimes I} \right)$, where $G$ is an unitary operator that get a purification $\left| 0 \right\rangle \left| 0 \right\rangle  \to \left| \Psi  \right\rangle$ by $G$ appling on  $\left| 0 \right\rangle \left| 0 \right\rangle $.  Thus, $
K'\left| M \right\rangle  = \frac{1}
{{M + 1}}\sum\limits_{p = 0}^M {\exp \left( { - \frac{1}
		{2}\left| {\vec x_M  - \vec x_p } \right|^2 } \right)} \left| p \right\rangle  = \frac{1}
{{M + 1}}\sum\limits_{p = 0}^M {\exp \left( { - \frac{1}
		{2}\left| {\vec x_*  - \vec x_p } \right|^2 } \right)} \left| p \right\rangle$ can be obtained.
In conclusion, covariance matrix $K$ and kernel function vector $\vec k_* $ can be prepared. Finally, the preparation of $\left| {\vec k_* } \right\rangle$ needs $O\left( {\frac{{MNt{\text{poly}}\log N\log \left( {\frac{t}
				{\varepsilon }} \right)}}
	{{\log \log \left( {\frac{t}
				{\varepsilon }} \right)\delta }} + T} \right)$. $O\left( T \right)$ always refer to the complexity of implement the blocking-encoding. Without loss of generality, $O\left( T \right) = O\left( {\log M} \right)$.

\section{Runtime analysis of the algorithm}
Let us start with discussing the time complexity of whole algorithm. As shown in TABLE 1, a detailed analysis of each step of this algorithm is depicted as follows. In A section, the state $\left| {\vec y} \right\rangle $ and $\left| {\vec k_* } \right\rangle $ can be generated in time $O\left( {{\text{poly}}\log M} \right)$ with the help of QRAM. Then, matrix exponentiation is performed in Step B1, which can be simulated within $\delta$ in time $
O\left( {\frac{{t{\text{poly}}\log M\log \left( {\frac{t}
				{\varepsilon }} \right)}}
	{{\log \log \left( {\frac{t}
				{\varepsilon }} \right)}}} \right)$. The phase estimation is performed in Step B2, this step by $O\left( {\frac{1}
{{t_0 }}} \right)$ in estimating eigenvalue. Generally, $\lambda _j  \geqslant \frac{1}
{\kappa }$ ($\kappa $ is the conditional number of covariance matrix $K$), taking $
t_0  = O\left( {\frac{\kappa }
	{{\delta '}}} \right)$ induces a final error of $\delta '$. Next, in Step B3, the runtime of controlled rotation is $O\left( {\frac{{\log \left( {\frac{1}
			{{\delta '}}} \right)\log ^2 \left( {\frac{\kappa }
			{{\delta '}}} \right)}}
{{\log \log \left( {\frac{1}
			{{\delta '}}} \right)}}} \right)$, which is relatively negligible
		compared to the time taken in Step B2. The inverse phase etimation has same time analysis with the phase estimation in Step B4. The Section C and Section D only utilize unitary operations, which is linear in the number of qubits, and the final measurement only account for a constant factor, so the runtime of these two sections is negligible. The aim of Section E is preparing covariance matrix $K$ and kernel function vector $\left| {\vec k_* } \right\rangle$. Through coherent state and block-encoding techniques, time complexity are $O\left( {\frac{{MNt{\text{poly}}\log N\log \left( {\frac{t}
				{\varepsilon }} \right)}}
	{{\log \log \left( {\frac{t}
				{\varepsilon }} \right)\delta }}} \right)$ and $O\left( {\frac{{MNt{\text{poly}}\log N\log \left( {\frac{t}
				{\varepsilon }} \right)}}
			{{\log \log \left( {\frac{t}
			{\varepsilon }} \right)\delta }} + {\log M}} \right)$ respectively. Without loss of generality, letting $O\left( N \right) = O\left( {\log M} \right)$. Therefore, the total runtime of getting predictive mean value and covariance value is $
\tilde O\left( {\frac{{tM{\text{poly}}\log M\log \left( {\frac{t}
{\varepsilon }} \right)}}
{\delta }} \right)$, where where the tilde indicates that we are neglecting more slowly growing terms. Compared with the runtime $O\left( {M^3 } \right)$ of classical computation, which achieves a quadratic acceleration.
\begin{table}
	\centering
	\caption{The time complexity of each step of whole algorithm }
	\begin{tabular}{cc}
		\hline
		\hline
		Step of Section   &     Time Complexity\\
		\hline
		Section A & $O\left( {\text{poly}\log M} \right)$
		\\
		Step B1 &   $O\left( {\frac{{t{\text{poly}}\log M\log \left( {\frac{t}{\varepsilon }} \right)}}{{\log \log \left( {\frac{t}{\varepsilon }} \right)}}} \right)$ \\
		Step B2 & $O\left( {\frac{{\delta '}}
			{k}} \right)$\\
		Step B3 & $O\left( {\frac{{\log \left( {\frac{1}
						{{\delta '}}} \right)\log ^2 \left( {\frac{\kappa }
						{{\delta '}}} \right)}}
			{{\log \log \left( {\frac{1}
						{{\delta '}}} \right)}}} \right)$\\
		Step B4 & $O\left( {\frac{{\delta '}}
			{k}} \right)$\\
		Sectin E & $O\left( {\frac{{MNt{\text{poly}}\log N\log \left( {\frac{t}
						{\varepsilon }} \right)}}
			{{\log \log \left( {\frac{t}
						{\varepsilon }} \right)\delta }} + {\log M}} \right)$\\
		\hline
		\hline
	\end{tabular}
\end{table}

\section{Conclusion}
Gaussian process regression makes sense in real-world applications, especially when problem involves extrapolating from large data sets. However, classical computers cost too much when data sets is large, thus quantum gaussian regression is proposed. In this work, firstly, we avoid the sparse Hamiltonian simulation and apply the LCU Hamiltonian simulation, making it possible to deal with the non-sparse matrix. Secondly, in order to get exact measurement value, that is the sign of results is not ambiguous, the original HHL algorithm \cite{HHL09} is improved, thus,  inner product can get a clear symbol. Another innovation point of this paper is the preparation of covariance matrix. In the existing papers on quantum gaussian process regression \cite{ZFF19}, authors have given the covariance matrix by default. In our paper, the covariance matrix is obtained by annihilation operator and generation operator and the kernel function vector is obtained by applying block coding. Finally, our algorithm has a quadratic acceleration compared with  classical counterpart.
We hope that our algorithm, especially, the key technologies used in our algorithm, the improved HHL algorithm and the matrix algorithm, will inspire more efficient quantum machine learning algorithms for application in a wider range of fields.

\begin{acknowledgments}
This work was supported by National Natural Science Foundation of China (Grant Nos. 61772134, 61976053 and 62006105), Fujian Province Natural Science Foundation (Grant No. 2018J01776) and Program for New Century Excellent Talents in Fujian Province University.
\end{acknowledgments}

\end{document}